      [1999/12/01 v1.4c Il Nuovo Cimento]
\documentclass[preprint]{cimento}

\usepackage{cite}
\usepackage{cancel}
\usepackage{amsmath}
\usepackage{graphicx}
\newcommand{\eq}[1]{Eq.~(\ref{#1})}
\newcommand{\ps}{p\hspace{-0.44em}/\hspace{0.06em}}
\newcommand{\Msusy}{M_{\textrm{SUSY}}}
\newcommand{\Bbar}{\,\overline{\!B}}
\newcommand{\gev}{\, \mathrm{GeV}}
\newcommand{\tev}{\, \mathrm{TeV}}
\newcommand{\bbs}{\ensuremath{B_s\!-\!\Bbar{}_s\,}}

\title{Chiral Enhancement in the MSSM -- An Overview}
\author{A.~Crivellin\from{ins:x}}
\instlist{\inst{ins:x} Albert Einstein Center for Fundamental Physics, Institute
  for Theoretical Physics,\\ University of Bern, CH-3012 Bern,
  Switzerland.}
\PACSes{14.65.Fy,14.80.Da,14.80.Ly}
\begin{document}

\maketitle

\begin{abstract}
In this article I review the origin and the effects of chirally enhanced loop-corrections in the MSSM based on Refs.~\cite{Crivellin,Crivellin:2008mq,Crivellin:2010ty}. Chiral enhancement is related to fermion-Higgs couplings (or self-energies when the Higgs field is replaced by its vev). I describe the resummation of these chirally-enhanced corrections to all orders in perturbation theory and the calculation of the effective fermion-Higgs and gaugino(higgsino)-fermion vertices. As an application a model with radiative flavor-violation is discussed which can solve the SUSY-CP and the SUSY-flavor problem while it is still capable of explaining the observed deviation from the SM in the \bbs mixing phase.
\end{abstract}

\section{Introduction}

In this article I summarize recent progress in the field of chirally enhanced self-energies in the MSSM done in collaboration with Jennifer Girrbach, Lars Hofer, Ulrich Nierste, Janusz Rosiek and Dominik Scherer. The discussion will skip the technical aspects and subtleties and instead focus on the essential features. The interested reader is referred to Refs.~\cite{Crivellin,Crivellin:2008mq,Crivellin:2010ty} for a detailed discussion.

In the standard model chirality violation is suppressed by small Yukawa couplings (except for the top quark). However, the MSSM does not necessarily possess the same suppression effects since the bottom Yukawa coupling can be big at large of $\tan\beta=v_u/v_d$ and also the trilinear $A$-terms do not necessarily respect the hierarchy of the Yukawa couplings. Thus, chirality-flipping self-energies can be enhanced either by a factor of $\tan\beta$ \cite{Hall:1993gn} or by a ratio $A^q_{ij}/(v Y^{q}_{ij})$. This enhancement can compensate for the loop suppression leading to corrections which are of the same order as the corresponding physical quantities, i.e. of order one. These large corrections must be taken into account to all orders in perturbation theory (Sec.~\ref{renormalization}) leading to effective fermion-gaugino and fermion-Higgs couplings (see Sec.~\ref{vertices}). 

It is even possible that light-fermion masses and the CKM elements are entirely due to radiative corrections involving the trilinear $A$-terms~\cite{Buchmuller:1982ye}. Such a model of radiative flavor-violation can solve the SUSY-CP \cite{Borzumati:1999sp} and the SUSY-flavor problem while still leading to interesting effects in flavor-observables (see Sec.~\ref{radiative}). 

\section{Self-energies}

One can decompose any fermion self-energy into chirality-flipping and
chirality-conserving parts in the following way:
\begin{equation}
\Sigma_{ji}^f(p) = \left( {\Sigma_{ji}^{f\,LR}(p^2) + \ps\Sigma
  _{ji}^{f\,RR}(p^2) } \right)P_R + \left( {\Sigma_{ji}^{f\,RL}(p^2) +
  \ps\Sigma_{ji}^{f\,LL}(p^2) } \right)P_L\,.
\label{self-energy-decomposition}
\end{equation}
Here $i$ and $j$ are flavor indices running from $1$ to $3$. Since the SUSY particles are known to be much heavier than the SM fermions it is possible to expand in the external momentum. 
For our purpose it is even sufficient to work in the limit $\cancel{p}=0$ in which only the chirality-flipping parts $\Sigma_{ji}^{f\,LR}(0)=\Sigma_{ij}^{f\,RL\star}(0)$ remain.
It is well known that in the MSSM these self-energies can be enhanced either by a factor $\tan\beta$
\cite{Hall:1993gn} or by a factor $A^{f}_{ij}/(Y^{f}_{ij}\Msusy)$
\cite{Crivellin:2008mq} which compensates for the loop-suppression and leads to order one corrections.  The chirality-changing part of the fermion self-energy (at $\cancel{p}=0$, involving sfermions and gauginos(higgsions)) can be written as
\begin{equation}
\Sigma_{ji}^{f\tilde \lambda \,LR} = \dfrac{-1}{16\pi^2}
\sum\limits_{s = 1}^6 \sum\limits_{I = 1}^N m_{\tilde \lambda_I}
\Gamma_{f_j \tilde f_s }^{\tilde \lambda_I L\star} \Gamma_{f_i \tilde f_s
}^{\tilde \lambda_I R} B_0 \left(m_{\tilde \lambda_I }^2,
m_{\tilde f_s }^2 \right).
\label{MSSM-self-energies}
\end{equation}
Here $\tilde{\lambda}$ stands for the SUSY fermions ($\tilde{g},
\tilde{\chi^0}, \tilde{\chi^\pm}$) and $N$ denotes their corresponding
number (2 for charginos, 4 for neutralinos and 8 for gluinos).  The
coupling coefficients $\Gamma^{\tilde\lambda_I L(R)}_{f_i \tilde f_s }$ and the loop
functions $B_0$ are defined in the appendix of Ref.~\cite{Crivellin}.  

The couplings $\Gamma^{\tilde \lambda_I L(R)}_{f_i \tilde f_s }$ in
\eq{MSSM-self-energies} involve the corresponding sfermion mixing
matrices $W^{f}$ which diagonalize the sfermion mass matrices: $W^{f\star}_{s^\prime s} (M_{\tilde f}^2)_{s^\prime t^\prime} W^{f}_{t^\prime t}=m_{\tilde f_s}^2\delta_{st} $. In the case of neutralino-quark-squark and chargino-quark-squark vertices they also depend on Yukawa couplings and CKM elements. 

An interesting feature of the self-energies in \eq{MSSM-self-energies} is that they are finite and that they
do not vanish in the limit of infinitely heavy SUSY masses. We refer
to this approximation in which only such non-decoupling terms for the self-energies are kept
as "the decoupling limit".  Note, however, that we do not integrate out the SUSY
particles but rather keep them as dynamical degrees of freedom.

Let us take a closer look at the quark self-energy with squarks and gluinos as virtual particles\footnote{The gluino contribution is the dominant one in the flavor-conserving case. In the presence of non-minimal sources of flavor-violation it is also usually the dominant one.}.
To leading order in $v/\Msusy$, the self-energy is proportional to one chirality flipping element
$\Delta_{jk}^{q\,LR}$ of the squark mass matrix:
\begin{eqnarray}
\Sigma_{fi}^{q \tilde{g}\,LR} &=& \dfrac{2\alpha_s}{3\pi}\, m_{\tilde
  g} \sum\limits_{j,k,j^\prime,f^\prime = 1}^3\; \sum\limits_{s,t =
  1}^6 \Lambda_{s\,fj}^{q\,LL}\; \Delta _{jk}^{q\,LR}\;
\Lambda_{t\,ki}^{q\,RR}\; C_0\! \left( m_{\tilde g}^2, m_{\tilde
  q_m^L}^2 ,m_{\tilde q_n^R }^2 \right).
\label{eq:gluinoSE}
 \end{eqnarray}
Here the off-diagonal elements of the sfermion mass matrices are given by
\begin{equation}
\begin{array}{l}
\Delta^{u\,LR}_{ij}=-v_u A^u_{ij}\;-\;v_d A^{\prime
  u}_{ij}\;-\;v_d\,\mu\, Y^{u_i}\, \delta_{ij}\,,\\
\Delta^{d,\ell\,LR}_{ij}=-v_d A^{d,\ell}_{ij}\;-\;v_u A^{\prime d,\ell}_{ij}\;-\;v_u\,
\mu\, Y^{d_i,\ell_i} \delta_{ij}\,,
\label{DeltaLR}
\end{array}
\end{equation}
and the matrices $\Lambda_{s\,ij}^{f\,LL,RR} \,=\, W^{f\star}_{i+3,s}\,W^{f}_{js}$ take into account the flavor changes due to bilinear terms. Note that in the decoupling limit $W^{f}_{st}$ depends only on the bilinear terms.

For equal SUSY masses we can give a simple approximate formula for the self-energy in \eq{eq:gluinoSE}:
\begin{eqnarray}
\Sigma_{fi}^{d \tilde{g}\,LR} &=& \dfrac{-1}{100}v_d\left( A^d_{fi}/\Msusy + Y^{d_i}\tan\beta \delta_{fi}  \right)\\
\Sigma_{fi}^{u \tilde{g}\,LR} &=& \dfrac{-1}{100}v_u\left( A^u_{fi}/\Msusy + Y^{u_i}\cot\beta \delta_{fi}  \right)
\label{SE_approx}
 \end{eqnarray}
Thus, generic $A^d$-terms which are of the order $M_{\rm{SUSY}}$ lead to self-energies which are approximatly $v_d/100$. The part of \eq{SE_approx} containing $Y^b$ is of the order of 2 GeV for $\tan\beta\approx50$. In the case of up-quarks, only the part of the self-energy proportional to $A^u$ can be important: it is of the order of 1.5 GeV for $A^u\approx M_{\rm{SUSY}}$.

According to \eq{DeltaLR} and \eq{SE_approx} the quark self-energy with a gluino as virtual particle can be divided into a part linear in a Yukawa coupling and a part linear in an $A$-term. Such a decomposition is possible for all self-energies (in the decoupling limit) because either a Yukawa coupling or a trilinear $A$-term is needed in order flip the chirality. Thus, we can also decompose the chargino self-energies (we do not consider the neutralino self-energy here, because it is usually subleading) in an analogous way. In the flavor changing case we also have to distinguish whether the flavor-change is due to a CKM element or not which is important when we consider later the CKM renormalization. Thus we decompose the down-quark self-energy as follows:
\begin{equation}\renewcommand{\arraystretch}{2.0}
\begin{array}{l}
\Sigma_{ii}^{d\,LR} = \Sigma_{ii\,\cancel{Y^{d_i}}}^{d\,LR}\,+\,v_u\,\,Y^{d_i} \epsilon_i^{d}\\
\Sigma_{fi}^{d\,LR} = \Sigma_{fi\,\cancel{CKM}}^{d\,LR}\,+\,m_{d_3} V^{(0)\star}_{3f}\epsilon_{FC}\delta_{i3}\;\;\rm{for}\;\; f\neq i 
\end{array}
\end{equation}
Here $\epsilon_i^{d}$ is the part of the flavor-conserving down-quark self-energy proportional to $Y^{d_i}$ divided by $v_u Y^{d_i}$, $\Sigma_{fi\,\cancel{CKM}}^{d\,LR}$ is the sum of all self-energies where the flavor-change is not due to CKM elements and $\epsilon^d_{FC}$ arises from the part of the chargino self-energy where the flavor change comes from a CKM element:
\begin{equation}
\epsilon^d_{FC} = \dfrac{-1}{16\pi^2}\,\mu\,
\dfrac{Y^{d_3}}{m_{d_3}} \sum\limits_{m,n = 1}^{3}
Y^{u_3\star}\,\Lambda_{m\,33}^{q\,LL}\,\Delta_{33}^{u\,LR\star}\,
\Lambda_{n\,33}^{u\,RR}\, C_0 \left( \left| \mu \right|^2 ,m_{\tilde
  q_m^L }^2 ,m_{\tilde u_{n}^R }^2 \right).\label{eq:epsFC}
\end{equation}
For the discussion of the effective Higgs vertices we also need a decomposition of
$\Sigma_{ji}^{f\,LR}$ into its holomorphic and non-holomorphic parts. Here non-holomorphic means that the loop-induced coupling is to the other Higgs doublet than the on involved in the Yukawa-term of the superpotential, i.e. for down-quarks the self-energy involves $v_u$ and for up-quarks the self-energy contains $v_d$. In the decoupling limit all enhanced holomorphic self-energies are proportional to $A$-terms and we denote the
the sum as $\Sigma_{ji\,A}^{f\,LR}$, while the
non-holomorphic part is denoted as $\Sigma_{ji}^{\prime f \,LR}$:
\begin{equation}
\Sigma_{ji}^{f\,LR} = \Sigma_{ji\,A}^{f\,LR} + \Sigma_{ji}^{\prime f\,LR}\,.  \label{HoloDeco}
\end{equation}

\section{Renormalization}
\label{renormalization}

Chirally-enhanced self-energies modify the relation between the bare
Yukawa couplings $Y^{q_i}\equiv Y^{q_i}$ and the corresponding physical fermion
masses $m_{f_i}$. For quarks we have the relation:
\begin{equation}
m_{q_i }  \;=\; v_q Y^{q_i}  \,+\, \Sigma_{ii}^{q\,LR},\hspace{1cm} (q=u,d) \label{mq-Yq}.
\end{equation}
\eq{mq-Yq} implicitly determines the bare Yukawa couplings
$Y^{q_i}$ for a given set of SUSY parameters.  
In the up-quark sector the enhanced terms in the self-energy
$\Sigma_{ii}^{u\,LR}$ are independent of $Y^{u_i}$. Therefore
\eq{mq-Yq} is easily solved for $Y^{u_i}$ and one finds
\begin{equation}
Y^{u_i} = \left(m_{u_i } - \Sigma_{ii}^{u\,LR}\right)/v_u .
\label{mu-Yu}
\end{equation}
In the down-quark sector we have terms proportional to one power of $Y^{d_i}$ at
most (in the decoupling limit) and by solving \eq{mq-Yq} we recover the well-known resummation formula \cite{Carena:1999py} with an extra correction due to the
$A$-terms \cite{Guasch:2003cv}\footnote{This equation can be directly transferred to the lepton sector by replacing fermion index $d$ for $\ell$, except for the vev.}:
\begin{equation}
Y^{d_i} = \frac{m_{d_i} - \Sigma_{ii\,\cancel{Y_i}}^{d\,LR}}{v_d\left( {1 + \tan\beta \epsilon_i^d } \right)}
\label{md-Yd}
\end{equation}

The flavor-changing self-energies $\Sigma^{q\,LR}_{fi}$ induce
wave-function rotations $\psi^{f\,L,R}_{i}\,\to\, U^{q\,L,R}_{ij}\,\psi^{q\,L,R}_j$
in flavor-space which have to be applied to all external fermion
fields. At the two-loop level $U^{q\,L}_{fi}$ is given by
\begin{equation} \renewcommand{\arraystretch}{2.0}
U^{q\,L}  \,=\, 
\left( {\begin{array}{*{20}c}
1- \frac{1}{2}\,\left|\sigma^q_{12}\right|^2 & \sigma^q_{12}\,+\,\frac{m_{q_1}}{m_{q_2}}\,\sigma^{q\star}_{21} &
    \sigma^q_{13}\,+\,\frac{m_{q_1}}{m_{q_3}}\,\sigma^{q\star}_{31} \\
    -\sigma^{q\star}_{12}\,-\,\frac{m_{q_1}}{m_{f_2}}\,\sigma^{q}_{21} & 1- \frac{1}{2} \,\left|\sigma^q_{12} \right|^2 &
    \sigma^q_{23}\,+\,\frac{m_{f_2}}{m_{q_3}}\,\sigma^{q\star}_{32} \\
    -\sigma^{q\star}_{13}\,-\,\frac{m_{q_1}}{m_{q_3}}\,\sigma^{q}_{31}+\sigma^{q\star}_{12}\,\sigma^{q\star}_{23} &
    -\sigma^{q\star}_{23}\,-\,\frac{m_{q_2}}{m_{q_3}}\,\sigma^{q}_{32} & 1 \\
\end{array}} \right),
\label{DeltaU}
\end{equation}
where we have neglected terms which are quadratic or of higher order
in small quark mass ratios and we have defined the abbreviation
$\sigma^q_{fi}=\Sigma^q_{fi}/m_{q_{\rm{max}(f,i)}}$.
The corresponding expressions for $U^{f\,R}$ is obtained from the
one for $U^{q\,L}$ by replacing $\sigma^q_{ji} \to
\sigma^{q\star}_{ij}$. The rotations in \eq{DeltaU} also renormalize the CKM matrix. The bare CKM matrix $V^{(0)}$, which arises because of the misalignment between the bare Yukawa couplings, is now determined through the physical one by
\begin{equation}
V^{(0)}  = U^{u\,L}\, V\, U^{d\,L\dag}.  
\label{CKM-0-ren}
\end{equation}
This equation can be solved analytically by exploiting the CKM hierarchy. First one calculates the effects of $\Sigma_{fi\,\cancel{CKM}}^{d\,LR}$ which lead to an additive change in the CKM elements. Then the self-energies containing CKM elements lead to a scaling of new CKM elements $\tilde V_{13},\tilde V_{23},\tilde V_{31},\tilde V_{32}$ elements by a factor $1/\left( {1 - \epsilon^d_{FC} } \right)$ (similar to \eq{md-Yd}).

\section{Effective Vertices}
\label{vertices}

\subsection{Higgs Vertices}

The effective Higgs vertices are most easily obtained in an effective field theory approach \cite{Isidori:2002qe} which is an excellent approximation to the full theory. In addition to the flavor-diagonal holomorphic couplings of quarks to the Higgs fields flavor-changing couplings to both Higgs doublets are induced via loops.
The resulting effective Yukawa-Lagrangian is that of a general 2HDM of type III
and is given (in the super-CKM basis) by:
\begin{equation} 
\begin{array}{l}
\mathcal{L}^{eff}_Y = \bar{Q}^a_{f\,L} \left[
  (Y^{d_i}\delta_{fi}+E^d_{fi}) \epsilon_{ab}H^b_d\,-\,E^{\prime
    d}_{fi} H^{a\star}_u \right]d_{i\,R}\\
-\bar{Q}^a_{f\,L} \left[ (Y^{u_i} \delta_{fi} +
  E^u_{fi})\epsilon_{ab}H^b_u \,+\, E^{\prime u}_{fi} H^{a\star}_d
  \right]d_{i\,R}\,.
  \end{array}
\end{equation}
Here $a$ and $b$ denote $SU(2)_L$\,-\,indices and $\epsilon_{ab}$ is the
two-dimensional antisymmetric tensor with $\epsilon_{12}=1$. The loop-induced couplings $E^{(\prime)q}$ are given by
\begin{equation}
   E^d_{ij}\,=\,\dfrac{\Sigma^{d\,LR}_{ij\,A}}{v_d}\,,\qquad
   E^{\prime d}_{ij}\,=\,\dfrac{\Sigma^{\prime d\,LR}_{ij}}{v_u}\,,\qquad
   E^u_{ij}\,=\,\dfrac{\Sigma^{u\,LR}_{ij\,A}}{v_u}\,,\qquad 
   E^{\prime u}_{ij}\,=\,\dfrac{\Sigma^{\prime u\,LR}_{ij}}{v_d}\,.
\end{equation}
Diagonalizing the effective quark mass matrices (after electroweak symmetry breaking) and decomposing the Higgs fields into their physical components leads to the following effective neutral Higgs couplings:
\begin{equation}
\begin{array}{l}
{\Gamma_{u_f u_i }^{H_k^0\,LR\,\rm{eff}} } = x_u^k\left( \frac{m_{u_i }}{v_u}
\delta_{fi} - \widetilde E_{fi}^{\prime u}\cot\beta \right) + x_d^{k\star}
\widetilde E_{fi}^{\prime u}\,,\\[0.2cm]
{\Gamma_{d_f d_i }^{H_k^0\,LR\,\rm{eff} } } = x_d^k \left( \frac{m_{d_i
}}{v_d} \delta_{fi} - \widetilde E_{fi}^{\prime d}\tan\beta \right) +
x_u^{k\star}\widetilde E_{fi}^{\prime d} \,,\\[0.2cm]
\end{array}
 \label{Higgs-vertices-decoupling}
\end{equation}
with
\begin{eqnarray}
\widetilde{E}^{\prime q}_{fi}=U_{jf}^{q\,L*} E^{\prime q}_{jk}
U_{ki}^{q\,R} \,\approx\, E_{fi}^{\prime q}\,-\, \Delta E_{fi}^{\prime
  q},\qquad\qquad\qquad\qquad\qquad\qquad\qquad\qquad\qquad\\
\Delta E^{\prime q} = \left( {\begin{array}{*{20}c} 0 &
    \sigma^q_{12} E_{22}^{\prime q}
    & \left(\sigma^q_{13}-\sigma^q_{12}\sigma^q_{23}\right)
    E^{\prime q}_{33}+\sigma^q_{12}E^{\prime q}_{23} \\
E_{22}^{\prime q}\sigma^q_{21} & 0 & \sigma^q_{23}E^{\prime q}_{33}
\\
E^{\prime q}_{33}\left(\sigma^q_{31}-\sigma^q_{32}\sigma^q_{21}\right)+E^{\prime q}_{32}\sigma^q_{21}\hspace{0.5cm} & E^{\prime q}_{33}\sigma^q_{32}
& 0
\end{array}} \right)\nonumber.
\label{Etilde}
\end{eqnarray}
The new term $\Delta E^{\prime q}$ is especially interesting: it contains a non-holomorphic flavor-conserving part which multiplies a flavor-changing holomorphic term. In this way the holomorphic $A$-terms can lead to flavor-changing neutral Higgs couplings. The origin of this term can be understood in the following way: Even though the couplings $E^{d}_{ij}$ are holomorphic, they lead to an additional rotation (if $E^{d}_{ij}$ is flavor non-diagonal) which is needed to diagonalize the effective quark mass matrix. These rotations then lead to off-diagonal neutral Higgs couplings even if $E^{\prime d}_{ij}$ is flavor conserving. This effect will allow us to explain the $B_s$ mixing in our model with radiative flavor-violation. 

\subsection{Gaugino(Higgsino)-Fermion-Vertices}

Effective gaugino(higgsino)-fermion-vertices which include the chirally-enhanced corrections are obtained by inserting the bare values for the Yukawa couplings and the CKM elements into the corresponding Feynman-rules and applying the wave-function rotations in \eq{DeltaU} to all external fermion fields. Since the genuine vertex corrections are not enhanced, these vertices then include all chirally enhanced effects.

\begin{figure}
\centering
\includegraphics[width=0.45\textwidth]{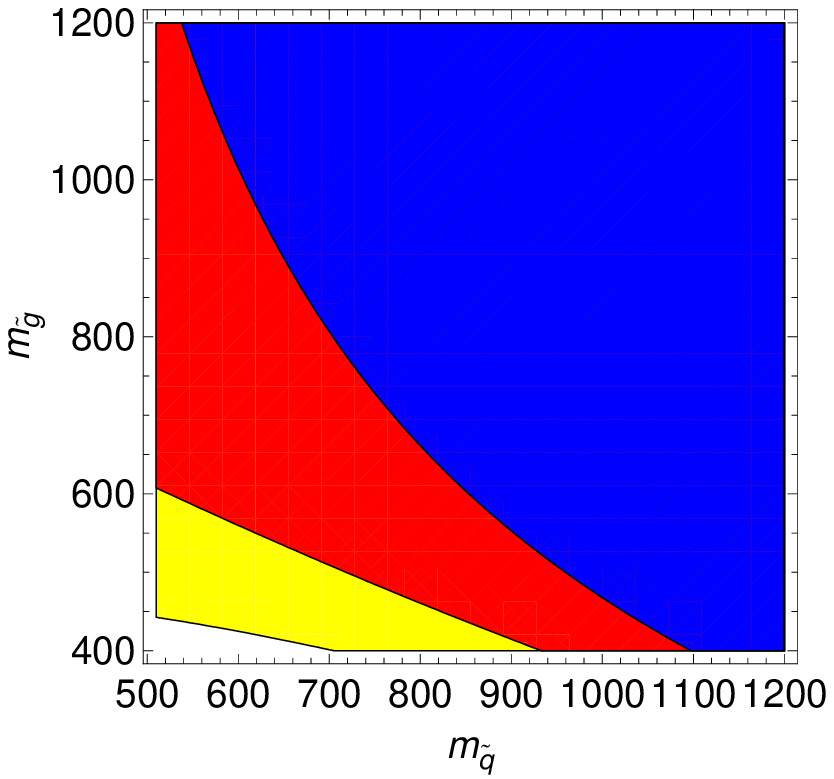}
\includegraphics[width=0.45\textwidth]{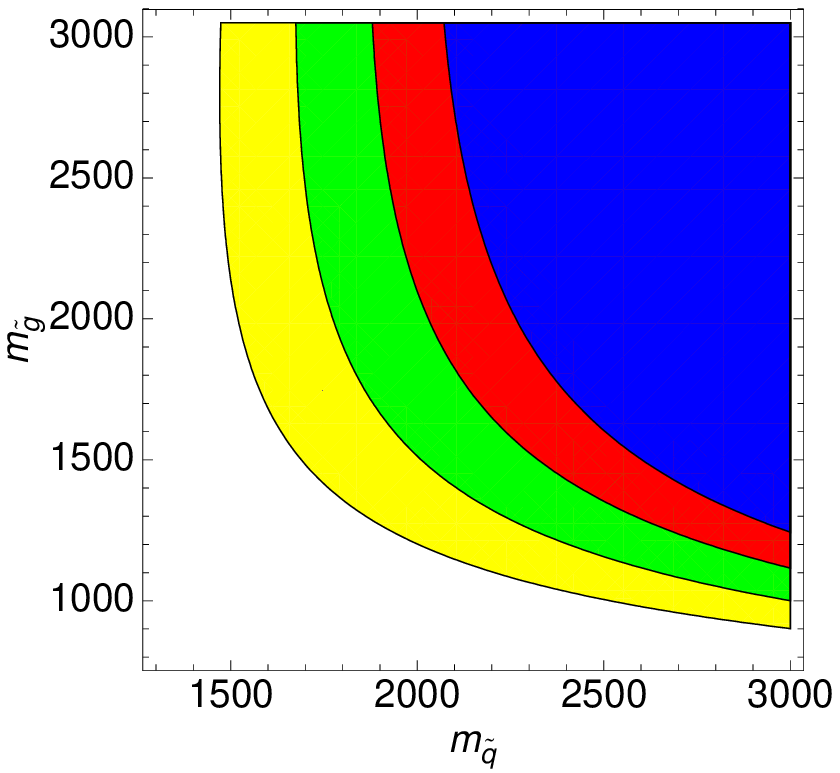}
\caption{{\small Left: Allowed regions in the $m_{\tilde g}-m_{\tilde q}$ plane. Constraint from $b\to s \gamma$ assuming that the CKM matrix is generated in the down sector. We demand that the gluino contributions should not exceed the SM one. Yellow(lightest): $\mu \tan\beta=30 \tev$, red: $\mu \tan\beta=0 \tev$ and blue: $\mu \tan\beta=-30 \tev$. \newline Right: Allowed regions in the $m_{\tilde g}-m_{\tilde q}$ plane. Constraints from Kaon mixing for different values of $M_2$ assuming that the CKM matrix is generated in the up sector. Yellow(lightest): $M_2=1000 \gev$, green: $M_2=750\gev$, red: $M_2=500\gev$ and blue: $M_2=250\gev$.}}
\label{constraints}
\end{figure}

\section{Radiative Flavor-Violation}
\label{radiative}

The smallness of the off-diagonal CKM elements and the Yukawa couplings of the first two generations suggests the idea that these quantities might be due to radiative corrections, i.e. they are zero at tree-level. Indeed, as we have see previously, the self-energies involving the trilinear $A$-terms lead to order one effects in the renormalization of the CKM elements and the light fermion masses and it is possible that they are generated by the self-energy radiative corrections \cite{Buchmuller:1982ye}. From \eq{SE_approx} we see that this is the case if the $A$-terms are of the same order as the other SUSY parameters.

However, the third generation fermion masses are too heavy to be loop generated (without unnaturally large values for the $A$-terms which would violate vacuum stability\cite{Casas:1995pd}) and the successful bottom -- tau (top -- bottom) Yukawa coupling unification in SU(5) (SO(10)) GUTs suggests to keep the third generation fermion masses. Thus we assume the following structure for the Yukawa couplings of the MSSM superpotential:
\begin{equation}
Y^{f}_{ij}  = Y^{f_{3}}\delta_{i3}\delta_{j3},\;\;\;V^{(0)}_{ij}  = \delta_{ij}\,.\label{eq:YukCKM}
\end{equation} 
This means that (in the language of \cite{D'Ambrosio:2002ex}) the global $[U(3)]^5$ flavor
symmetry of the gauge sector is broken down to $[U(2)]^5 \times [U(1)]^2$ by the Yukawa couplings of the third generation. Here the five $U(2)$ factors correspond to rotations of the
left-handed doublets and the right-handed singlets of the first two generation fermions in flavor space, respectively. 

Let us first consider the quark sector. Here we demand that the light quark masses and the off-diagonal CKM elements are generated by gluino self-energies. Regarding only the first two generations, no direction in flavor space is singled out by the Yukawa term in the superpotential and the Cabbibo angle is generated by a misalignment between $A^u$ and $A^d$\footnote{This also implies that the quark-squark gluino vertex is flavor-diagonal for transitions between the first two generations in the super-CKM basis.}. Regarding the third generation, the situation is different because their non-zero Yukawa couplings fix the quark-field rotations involving the third generation. Thus $V_{ub,cb,ts,td}$ are generated by a misalignment between the $A^u$, $A^d$ and the third generation Yukawa couplings. We will consider the two limiting cases in which the CKM elements arise only from a mismatch between ($A^d$) $A^u$ and ($Y^{d}$) $Y^{d}$ which we call CKM generation in the down (up) sector for obvious reasons. This means we require 
\begin{eqnarray}
   \Sigma_{23}^{d\;LR} = m_{b}V_{cb} \,\approx\,  -m_{b}V_{ts}^*\,, \qquad \Sigma_{13}^{d\;LR} = m_{b}V_{ub}\,,\;\\
    {\rm or}\;\;  \Sigma_{23}^{u\;LR} = -m_{t}V_{cb} \,\approx\, m_{t}V_{ts}^* \,,\qquad   \Sigma_{13}^{u\;LR} = m_{t}V_{td}^\star\,.
\label{CKMdown}
\end{eqnarray}

\begin{figure}
\includegraphics[width=0.47\textwidth]{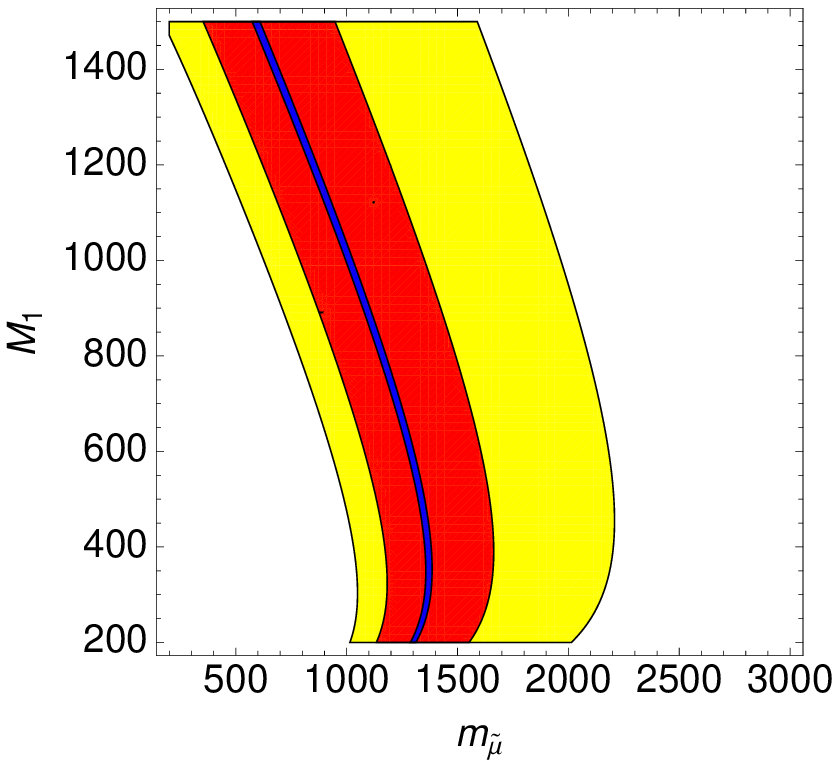}
\includegraphics[width=0.45\textwidth]{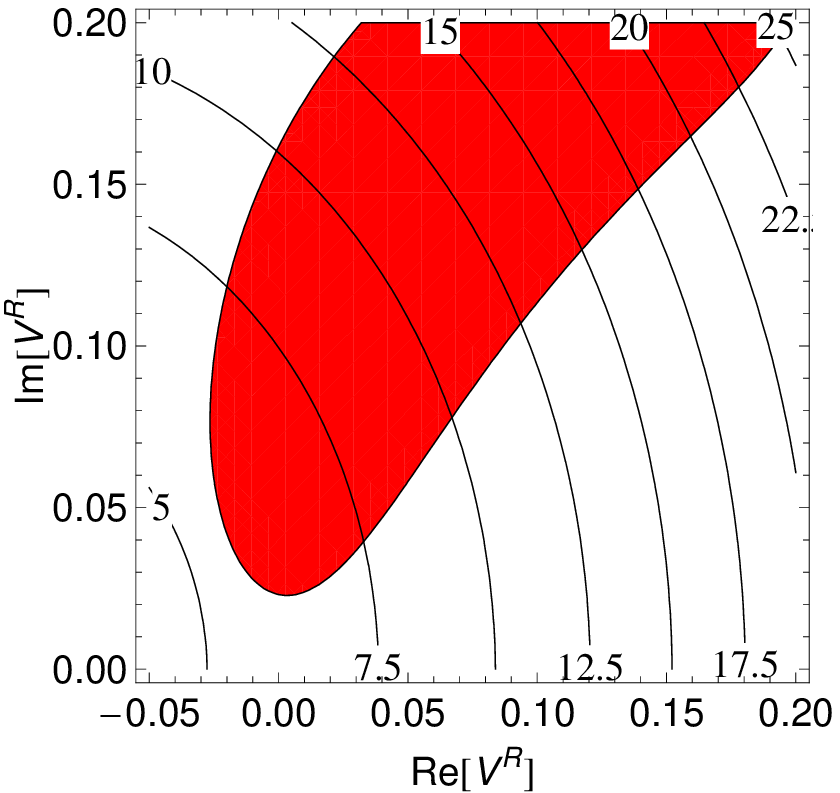}
\caption{{\small Left: Allowed region in the $M_1$-$m_{\tilde \mu}$  plane
  assuming that the muon Yukawa coupling is generated radiatively by
  $v_d A^\ell_{22}$ and/or $v_u A^{\ell\prime}_{22}$.  Here $m_{\tilde \mu}$
  is the lighter smuon mass. Yellow (lightest): $a_{\mu}\pm2\sigma$,
  red: $a_{\mu}\pm1\sigma$, blue (darkest): $a_{\mu}$.
  \newline Right: correlations between $B_s\to \mu^+\mu^-$ and \bbs mixing for $\epsilon_b=0.0075$, $m_H=400\rm{GeV}$ and $\tan\beta=12$ in the complex $V_R=\Sigma_{23}^{d\;RL}/m_{d_3}$ plane. Red: Allowed region from \bbs mixing (95\% confidence level). The contour-lines show $\rm{Br}[B_s\to \mu^+\mu^-]\times10^9$.}
  \label{muon_bs}}
\end{figure}

If the CKM matrix is generated in the down-sector the most stringent constraint stems from an enhancement of $b\to s \gamma$ due to the off-diagonal element $\Delta_{23}^{d\;LR}$ in the squark mass matrix. The resulting bounds on the squark and gluino mass are shown in the left plot of Fig.~\ref{constraints}.
In addition flavor-changing neutral Higgs coupling are induced according to \eq{Higgs-vertices-decoupling} which gives an additional contribution to $B_s\to\mu^+\mu^-$. Also $B_s$ mixing can be affected but because it is protected by a Peccei-Quinn symmetry a double Higgs penguins contributes only if also $\Sigma_{23}^{d\;RL}$ is non-zero (see right plot in Fig.~\ref{muon_bs}).

\begin{figure}
\centering
\includegraphics[width=0.75\textwidth]{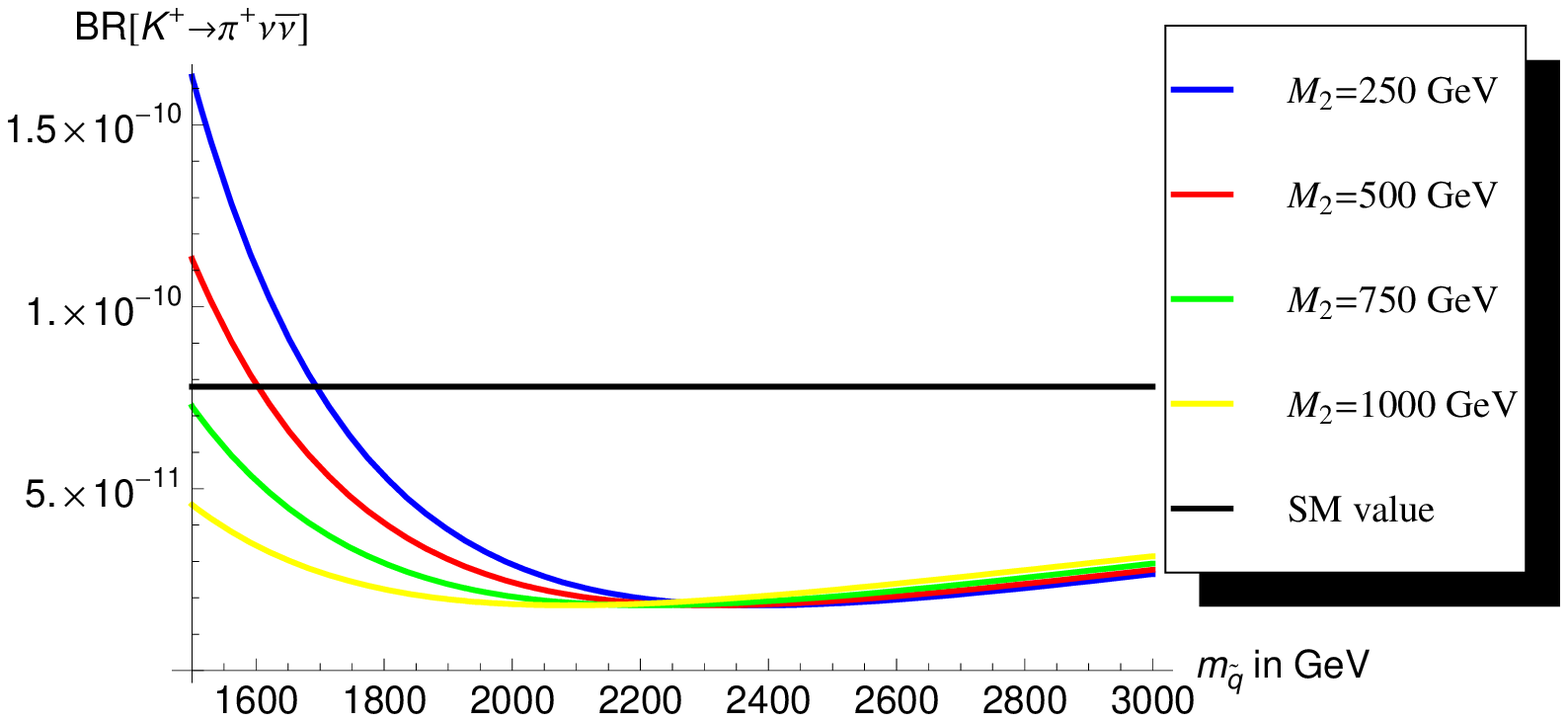}\\
\includegraphics[width=0.75\textwidth]{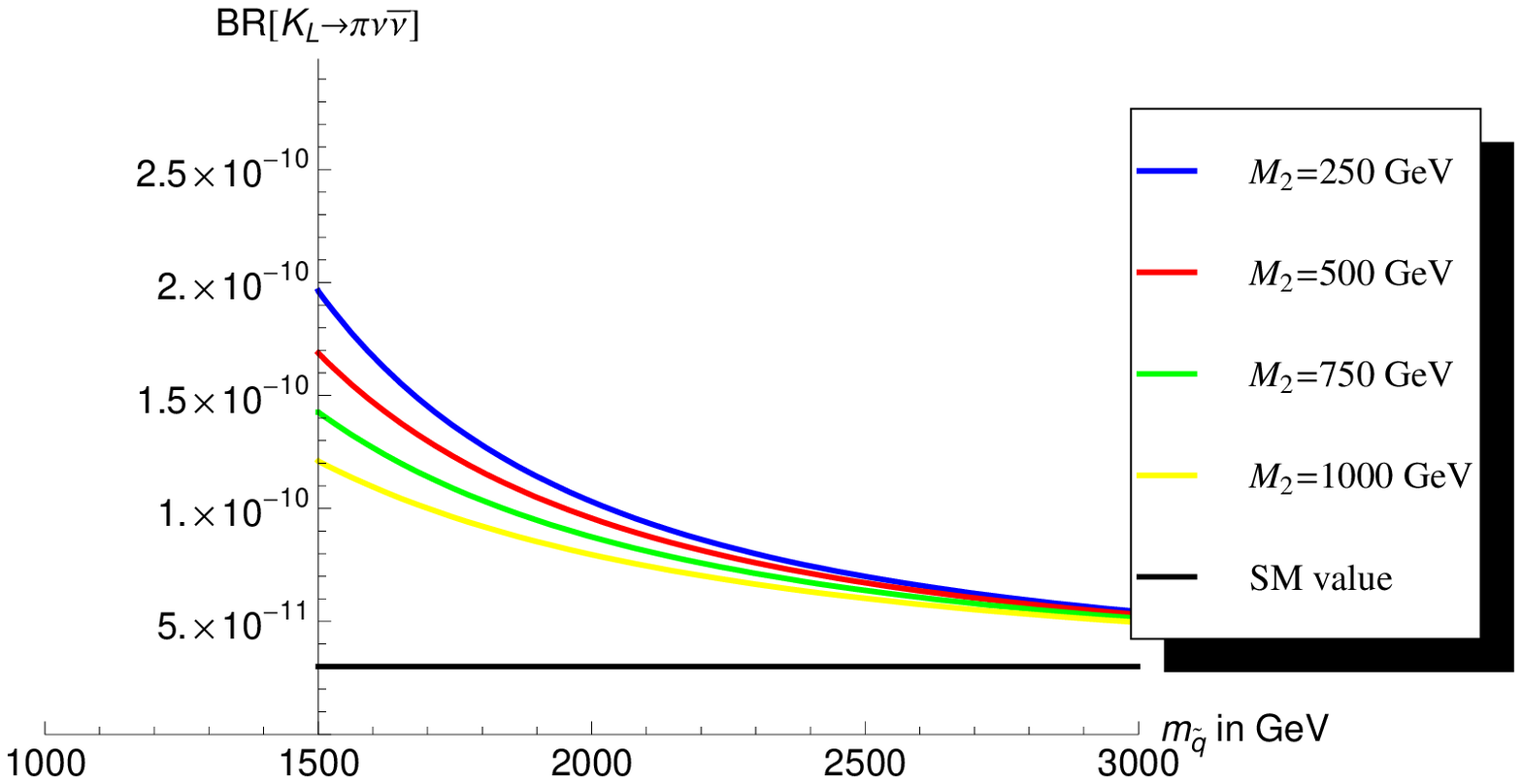}
\caption{{\small Predicted branching ratio for the rare Kaon decay $K_L\to\pi^0\nu\overline{\nu}$ (left) and $K^+\to\pi^+\nu\overline{\nu}$ (right) assuming that the CKM matrix is generated in the up-sector for $m_{\tilde{q}}=m_{\tilde{g}}$. 
\label{Kplustopinunu}}}
\end{figure}

In the case of CKM matrix generation in the up-sector,
the most stringent constraints stem from $\epsilon_K$ (see right plot of Fig.~\ref{constraints}) which receives additional contributions via a chargino box diagram involving the double mass insertion $\delta_{23}^{u\;LR}\delta_{13}^{u\;LR}$. At the same time the rare Kaon decays $K^+\to\pi^+\nu\overline{\nu}$ and $K_L\to\pi^0\nu\overline{\nu}$ receive sizable corrections (see Fig.~\ref{Kplustopinunu}) which is very interesting for NA62.

Radiative mass generation is also possible in the lepton sector. Here the anomalous magnetic moment of the muon probes the soft muon Yukawa coupling because it gets an additive contribution which depends only on the SUSY scale. If one demands that SM contribution plus the supersymmetric one is within the 2$\sigma$ region of the experimental measurement the smuon mass must lie between $600~\rm{GeV}$ and $2200~\rm{GeV}$ for
$M_1<1\rm{TeV}$ if its Yukawa coupling is loop-generated (see left plot of Fig.~\ref{muon_bs}). 
If a smuon is found to be lighter, the observed muon mass cannot entirely stem from the soft
SUSY-breaking sector and consequently the muon
must have a nonzero Yukawa coupling $y_\mu$ in the superpotential.

\section{Conclusions}

In the MSSM self-energies can be chirally enhanced by a factor $\tan\beta$ or by a factor $A^q_{ij}/(v Y^{q}_{ij})$ which can compensate for the loop-suppression. This leads to order one corrections which must be taken into account to all orders in perturbation theory. This goal can be achieved by using effective vertices which include these corrections.
The trilinear $A$-terms can even entirely generate the light fermion masses and the off-diagonal CKM elements via radiative corrections. Such a model of radiative flavor violation can both solve the SUSY CP problem and is consistent with FCNC constraints for SUSY masses of the order of 1 TeV. In addition in the case of CKM generation in the down-sector $B_s\to \mu^+\mu^-$ and \bbs mixing receive additional contributions via Higgs penguins which and can even generate a sizable phase in $B_s$ mixing. In the case of CKM generation in the up sector the branching ratio of the rare Kaon decays $K^+\to\pi^+\nu\overline{\nu}$ and $K_L\to\pi^0\nu\overline{\nu}$ can be enhanced compared to the SM prediction.

\acknowledgments

I thank the organizers, especially Gino Isidori, for the invitation to the "La Thuille conference". This work is supported by the Swiss National Foundation. I am grateful to Lars Hofer and Lorenzo Mercolli for proofreading the manuscript. The Albert Einstein Center for Fundamental
Physics is supported by the ``Innovations- und Kooperationsprojekt
C-13 of the Schweizerische Universit\"atskonferenz SUK/CRUS''.

\end{document}